Gold Bipyramids as a Promising Alternative to Gold Nanorods for Analytical and Biomedical Applications

Andrey M. Burov[1], Sergey V. Zarkov[1,2], Arina V. Drozd[1] , Igor V. Borisov[1], Elena G. Zavyalova[3], Nikolai G. Khlebtsov[1,4*]

*[1]Institute of Biochemistry and Physiology of Plants and Microorganisms, "Saratov Scientific Centre of the Russian Academy of Sciences," 13 Entuziastov Pr., Saratov 410049, Russia*

*[2]Institute of Precision Mechanics and Control, "Saratov Scientific Centre of the Russian Academy of Sciences," 24 Rabochaya Str., Saratov 410028, Russia*

*[3]Lomonosov Moscow State University, Faculty of Chemistry, Leninskie Gory 1-3, Moscow 119991, Russia*

*[4]Saratov State University, 83 Astrakhanskaya Str., Saratov 410012, Russia*

* E-mail: khlebtsov@ibppm.ru Cell phone:: +7 (905)3256536

**Graphical abstract**

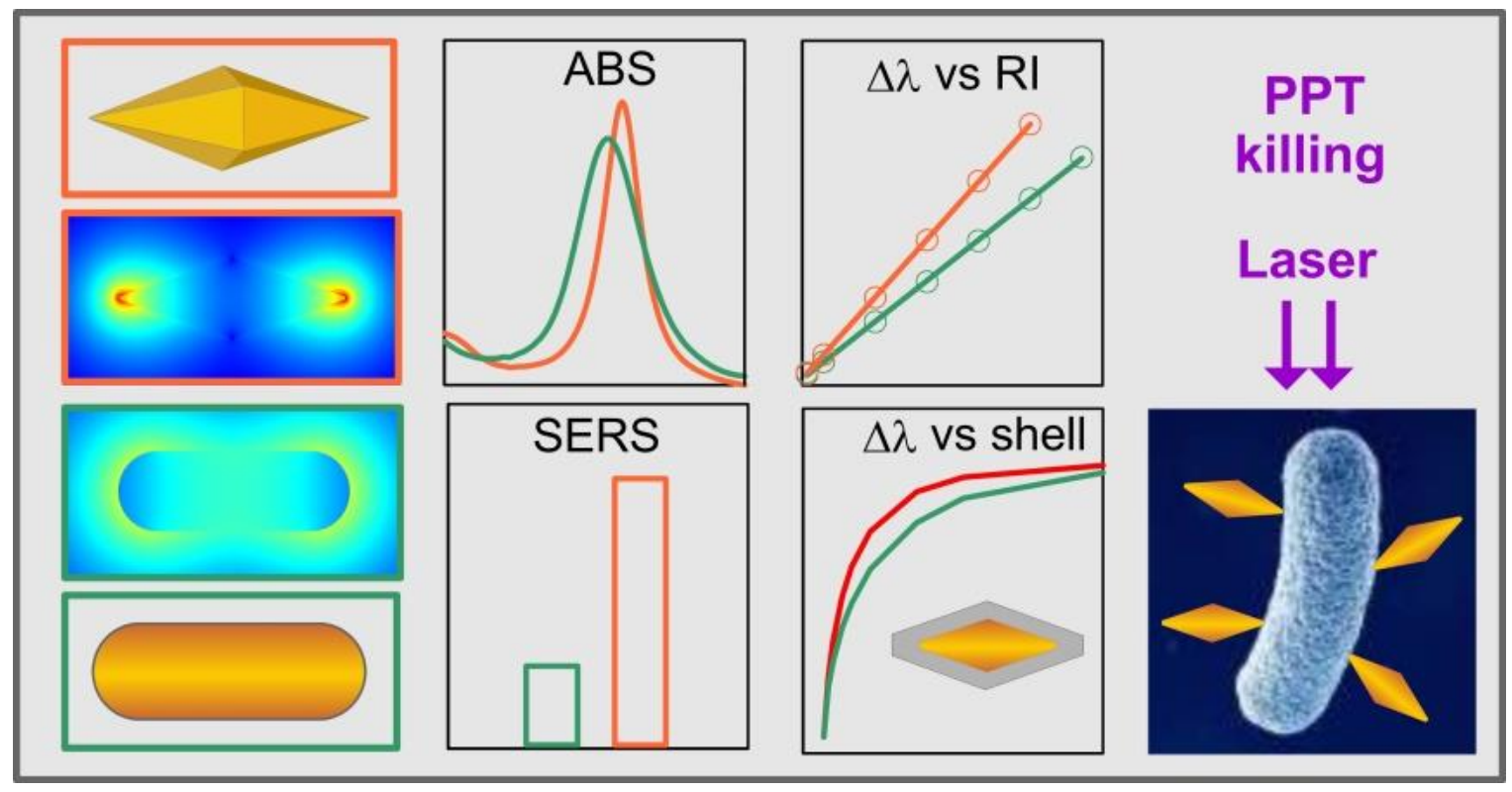

**Abstract**

Pentagonal gold bipyramids with dimensions of 75x25 nm and a longitudinal plasmon resonance (PR) at 753 nm are synthesized. For comparison, gold nanorods with a diameter of 20 nm, lengths ranging from 95 to 50 nm, and longitudinal PR from 945 to 644 nm were synthesized by chemical etching. The samples were characterized by UV-vis spectroscopy and transmission electron microscopy (TEM). It is shown that the absorption spectral quality factor of the bipyramids is significantly higher than that of the nanorods. To compare the nanoparticles as platforms for surface-enhanced Raman scattering (SERS), their surface was functionalized with thiolated nitrobenzene molecules (NBT). It is demonstrated that the SERS enhancement factor for the bipyramids is approximately three times higher than that for the nanorods. The red shift of the bipyramids' PR after functionalization with NBT molecules is also about three times greater than for nanorods with the same PR. This agrees with the theoretical estimate of the bipyramids' PR shift being more sensitive to variations in the refractive index of the external medium or the dielectric shell thickness than that of gold nanospheres and nanorods. The high efficiency of the bipyramids as thermosensitizers for converting laser radiation into heat in photothermal therapy is experimentally and theoretically demonstrated. Effective photothermal killing of E. coli was shown upon irradiation with a laser at the plasmon resonance wavelength using nanobipyramids or nanorods.

## 1. Introduction

Anisotropic metal nanoparticles [1-5] are actively researched and utilized due to their flexible tunability of the plasmon resonance (PR) across a broad UV-vis-NIR range [6], significantly larger scattering and absorption cross-sections [7], and enhanced local field effects [8] compared to plasmonic nanospheres. Modern methods of colloidal chemical synthesis [9] enable the production of a wide variety of anisotropic nanoparticles, including gold-silver nanocages [10], silver nanocubes [11], gold nanostars [12], triangular nanoprisms [13], and other flat particles [14], gold-silver nanorods [15], as well as particles with field enhancement in gaps (gap-enhanced Raman tags) [16]. However, the most well-known and widely used are conventional gold nanorods [17] with approximately hemispherical ends and a cylindrical or pentagonal cross-section. To date, the synthesis and application protocols for gold nanorods are well-established [18-21], which explains the consistently high annual number of publications on these nanoparticles (according to SCOPUS, over 180 articles in 2025 alone have titles containing the phrase "gold nanorods").

Compared to nanorods, pentagonal gold nanobipyramids (BPs) are far less studied [22], despite having similar plasmon resonance tunability [23] (with an aspect ratio change from 2 to 6, the plasmon resonance wavelength shifts from approximately 600 to 1100 nm) and thermoplasmonic properties [24-26]. Regarding the differences, BPs possess five uniformly distributed twinning planes at the vertices and sharp tips [27], whereas nanorods are single crystals with roughly hemispherical ends. These morphological differences result in distinct local plasmonic properties, higher resonance sensitivity to the dielectric environment [28-30], and higher quality factors of the plasmonic absorption and scattering peaks in BPs. Despite these apparent advantages, BPs remain, in our opinion, undeservedly less popular due to their non-trivial synthesis [31-36] and functionalization [37]. Furthermore, simulating the optical properties of BPs (especially when coated with functional molecules) requires numerical methods [38, 39], as until very recently there were no simple and sufficiently accurate analytical models for plasmonic nanoparticles with a dielectric coating [40].

This work presents a comparative theoretical and experimental study of the UV-vis extinction spectra, Surface-Enhanced Raman Scattering (SERS), refractive index sensitivity (RIS), thermoplasmonic properties of gold BPs and nanorods with similar plasmon resonance (PR) wavelengths, and plasmonic photothermal killing of bacteria in water suspensions. In contrast to published similar studies (see, e.g., [41, 42]), we compared the normalized quantities that are proportional to the SERS intensity per adsorbed molecule. For comparison of thermoplasmonic experiments, we introduced the normalized absorption efficiency that is proportional to the absorbed power per unit volume at a constant mass-volume concentration of particles. Finally, our analysis of analytical sensitivity includes two different options: (1) the usual sensitivity to the refractive index of the surrounding medium; (2) an original analysis of the analytical response to an adsorbed layer of analytes with different nanometer thicknesses.

Using the chemical etching method [43], we synthesized a set of nanorod samples with PR ranging from 600 to 950 nm, including a sample with a PR close to that of the BP sample. To calculate the plasmonic properties of the nanoparticles, we used a previously described 2.5D finite element method [44, 45] implemented in the commercial software package COMSOL Multiphysics 5.1 (Wave Optics module), applicable to axially symmetric particles and. Also, we used our recently developed analytical method [40].

We demonstrate that BP properties, including the quality factor of extinction spectra, SERS signal enhancement, the analytical sensitivity of the PR wavelength to the local dielectric environment, and the efficiency of light-to-heat conversion, surpass those of gold nanorods with comparable dimensions and PR wavelengths. *In vitro* experiments also demonstrate effective photothermal killing of E. coli DH5α bacteria under laser irradiation at a wavelength near the PR.

## 2. Methods

### *2.1. Theoretical methods*

In the 2.5D formalism [44, 45], the incident and scattered fields are expanded in cylindrical harmonics, and the response of each harmonic is computed separately, leading to significant savings in computational resources in terms of memory and processing time [46]. Compared to conventional 3D modeling, the 2.5D formalism substantially reduces the computational time per spectral data point and enables averaging cross-sectional spectra over random particle orientations with acceptable computational costs. The dielectric function of gold was calculated using a spline interpolation [47] of the data from [48]. In all calculations presented hereafter, the correction of the bulk dielectric function for the finite particle size was performed using the volume-equivalent sphere radius as the effective electron surface scattering length and the scattering constant $A_s = 0.33$ [49, 50]. The refractive index of the shell, $n_2 = \sqrt{\varepsilon_2} = 1.5$, was chosen to be close to that of many biopolymers and typical stabilizers, such as cetyltrimethylammonium bromide (CTAB) or cetyltrimethylammonium chloride (CTAC). In all calculations, the dielectric function of water [45] $\varepsilon_m(\lambda) = n_m^2(\lambda)$ was used as the external medium dielectric function. The thickness of the CTAB stabilizer shell was set to 3 nm based on the data from [51].

The employed analytical method MEM+DEM [40] is based on a combination of the modal expansion method (MEM) [52] and the dipole equivalence method (DEM) [53]. Hereafter, for brevity, we will refer to the MEM+DEM method simply as MEM.

The primary theoretical models considered were pentagonal bipyramids with tip rounding radii specified in [52] and cylindrical nanorods with hemispherical ends. We also employed an inscribed bicone as a bipyramid model, following the approach in [39].

### *2.2. Experimental methods*

The following reagents were used in this study: hydrogen tetrachloroaurate(III) trihydrate ($HAuCl_4 \cdot 3H_2O$); cetyltrimethylammonium bromide (CTAB, 98%); cetyltrimethylammonium chloride (CTAC, 25% solution in water); sodium borohydride ($NaBH_4$, 99%); silver nitrate ($AgNO_3$, 99%); hydrochloric acid (concentrated HCl); L-ascorbic acid (AA, 99.9%); citric acid (CA, 99%); 4-nitrothiophenol (NBT, 97%); hydroquinone (HQ, 99%). Ultra-pure deionized water obtained using a Milli-Q Integral 5 system was used in all experimental syntheses.

Extinction spectra were measured using a Specord 300 spectrophotometer (Analytik Jena, Germany). The Raman spectra of composite labels were recorded using a combined setup comprising a Leica MD 2500 microscope, a SeekerPro785 spectrometer from Ocean Optics (USA), and a 785 nm laser operating at 30 mW. The surface potentials (charges) of the nanoparticles were measured using a Zetasizer NanoSeries NT dynamic light scattering spectrometer (Malvern Instruments, UK). Nanoparticle images were obtained with a Libra-120 transmission electron microscope (TEM) (Carl Zeiss, Jena, Germany) at the "Symbiosis" Center for the Collective Use of Research Equipment (IBPPM RAS). A diode laser with a wavelength of 795-810 nm, fiber optic output, and a maximum continuous output power of 2 W (Opto Power Corp., USA) was used to heat nanoparticle suspensions. The temperature in the cuvette was measured using a Guide PC210 thermal imager (China).

Gold nanobipyramids were obtained by overgrowth of polycrystalline seeds in a CTAB medium [32]. In the first, preliminary stage, gold thermal seeds were synthesized. For this, 10 ml of 0.25 mM $HAuCl_4 \times 3H_2O$ were reduced with 0.25 ml of 25 mM sodium borohydride under vigorous stirring in a 20 ml flask, in an aqueous solution containing 50 mM CTAC and 5 mM citric acid. The reaction mixture changed color from yellow to brown. After 2 minutes, the solution was transferred to a silicone bath at 80°C and stirred for 90 minutes. During the reaction, the solution turned a deep red. In the second stage, the obtained seeds, cooled to room temperature, were added under vigorous stirring to a solution containing 50 ml of 100 mM

CTAB, 2.5 ml of 10 mM $HAuCl_4 \cdot 3H_2O$, 500 µl of 10 mM silver nitrate, 1 ml of 1 M hydrochloric acid, and 400 µl of 100 mM ascorbic acid. The mixture was kept at 30°C for about 2 hours. The resulting bipyramids were centrifuged several times (10,000 rpm) and redissolved in 1 mM CTAC. The synthesis yielded gold bipyramids with an average longitudinal size of 75 ± 3 nm, a transverse size of 25 ± 2 nm, and a plasmon resonance maximum around 753 nm.

Gold nanorods were obtained using a modified protocol [54] that enabled the synthesis of thin, long rods. Gold seeds were prepared by adding an aqueous sodium borohydride solution (10 mM, 0.6 ml) to an aqueous solution containing CTAB (0.1 M, 10 ml) and $HAuCl_4$ (10 mM, 0.25 ml). Silver nitrate (3.5 ml, 0.1 M) was added to a solution of $HAuCl_4$ (500 ml, 0.5 mM) in 0.1 M CTAB, followed by the addition of an aqueous hydroquinone solution (25 ml, 0.1 M). The resulting mixture was stirred manually until it became clear. Then, 8 ml of the seeds were added, the mixture was stirred, and left overnight at 30°C without stirring. A significant drawback of the protocol [54] is the high impurity particle content (10–15%). In this work, we used a modified version [55] with a purification step: the colloid was centrifuged at 10,000 g for 20 minutes, and the pellet was redissolved in 30 ml of 200 mM CTAC and left undisturbed for 1 hour. The nanorods aggregated at the bottom and the walls of the tube, forming a brown film. The supernatant, containing particles of other shapes, was removed. The nanorods were redissolved in 50 mM CTAB to a concentration corresponding to an optical density at the longitudinal resonance at 945 nm of 2, measured in a 2 mm pathlength cuvette. To completely remove impurity particles, the purification procedure was repeated twice. TEM analysis of the samples revealed the presence of thin nanorods with an average thickness of about 20 nm and a length of about 95 nm. A distinctive feature of our protocol is the practically zero percentage of non-target particles (0.5%) and the high quality of the extinction spectrum, as measured by the extinction ratio at the longitudinal and transverse PR wavelengths (>6).

To obtain AuNR samples with different longitudinal plasmon resonance (LPR) wavelengths, 50 ml of AuNR colloid in 50 mM CTAB was titrated with a 2 mM $HAuCl_4$ solution (adding 20 µl aliquots every 10 minutes). Before each addition, the extinction spectrum was measured, and if the desired LPR wavelength was reached, the required amount of sample (typically 3 ml) was taken from the colloid.

For the functionalization of gold nanorods and bipyramids with NBT molecules, 10 µl of a 2 mM ethanolic NBT solution was added to 1.5 ml of the colloid. The mixture was incubated for 30 minutes. The nanoparticles were then centrifuged at 10,000 rpm for 5 minutes and resuspended in 2 ml of 10 mM CTAC.

For *in vitro* photothermal experiments, a 24-hour bacterial culture of *E. coli* DH5α (from the collection of IBFPM RAS) with an optical density $A_{600} = 1$ in a 1-cm cuvette and gold nanorod and nanobipyramid suspensions with an optical density of $A_{800} = 5$ were used. A 0.01 M PBS buffer with pH=7.5 was used as the dilution medium. For the Alamar Blue assay [56], a resazurin solution at 0.55 mg/mL was used. Fluorescence spectra were recorded on a Cary Eclipse spectrofluorimeter (Agilent) over 590-610 nm, with a slit width of 5 nm and an excitation wavelength of 530 nm.

## 3. Results and discussion

### *3.1. Extinction spectra and geometrical parameters of nanoparticles*

In Figure 1A, solid lines represent the extinction spectra of four nanorod samples with LSPR at 945, 844, 735, and 644 nm (curves 1–4), as well as the spectrum of gold bipyramids (5); the inset shows a TEM image of the bipyramids. Dashed lines show the spectra of the same samples after functionalization with the Raman reporter molecule NBT. Panels B–E show TEM images of the initial 944 nm nanorods and of samples obtained through chemical length etching with virtually no change in thickness. The geometric parameters of the particles, based on statistical analysis of TEM images, are given in Table 1.

As shown in Table 1, during the etching process, the length of the nanorods decreases from 95 nm to 45 nm, and the aspect ratio decreases on average from approximately 4.85 to 2.4, while the nanorod diameter remains practically constant at 19±0.8 nm. Accordingly, the plasmon resonance wavelength shifts from the infrared (945 nm) to the visible (640 nm) region. The spectra in Fig. 1 were measured at approximately equal molar concentrations of gold (0.1 mM), as determined using the method described in [57].

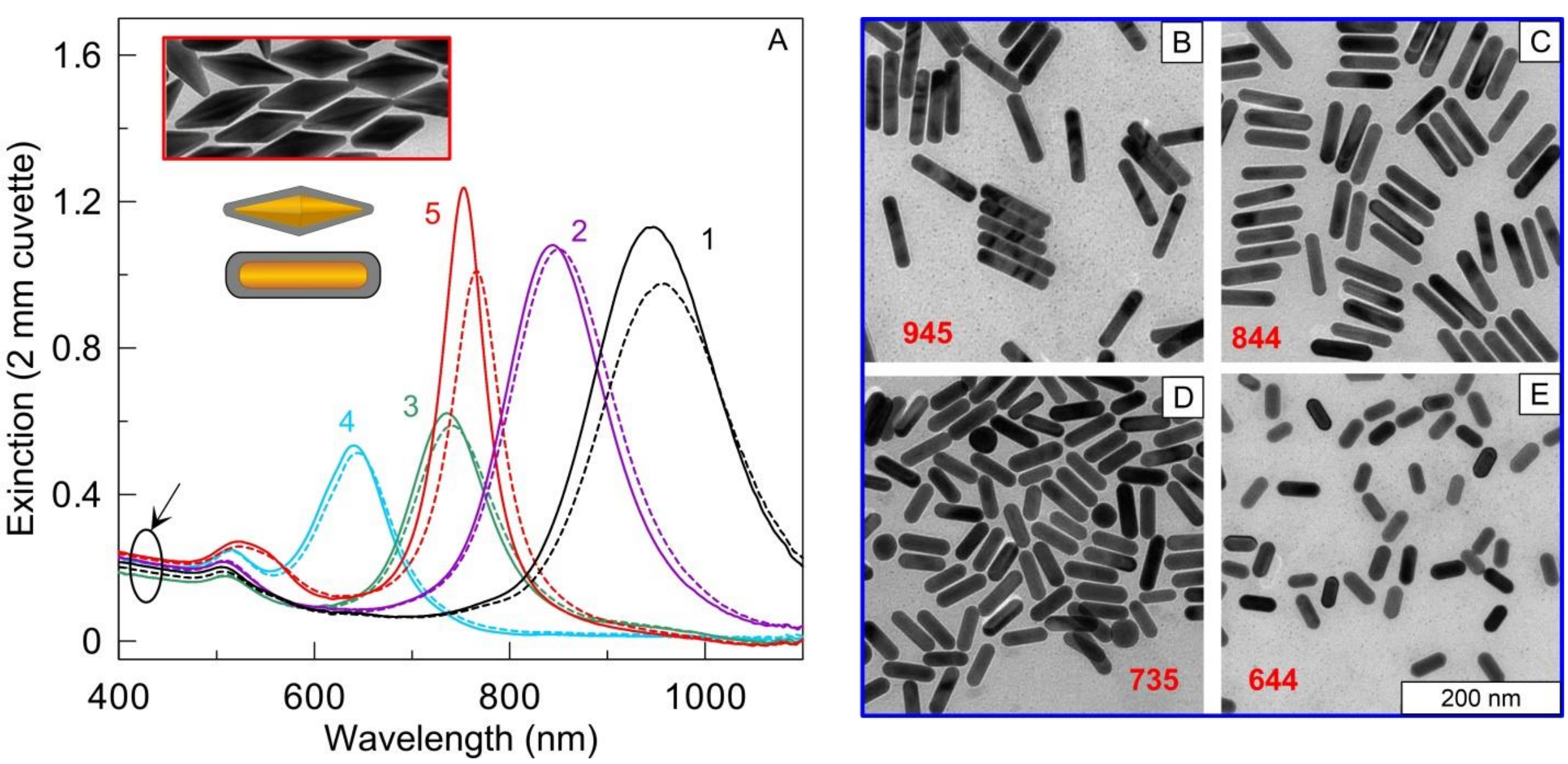

**Fig. 1**. Extinction spectra (measured in a 2 mm cuvette, panel A) and TEM images (panels B-E) of gold nanorods with plasmon resonances at 945 (1), 844 (2), 735 (3), and 644 (4) nm, and of gold bipyramids (inset in panel A) with a plasmon resonance at 753 nm (5). The arrow indicates the approximate equal extinction of all samples at 400 nm, corresponding to equal molar (or mass/volume) concentrations of gold. The inset shows a TEM image of the bipyramids. The dashed curves show the extinction spectra of the composite particles after functionalization with NBT molecules (AuNR@NBT, AuBP@NBT). Panel E shows a 200 nm scale bar applicable to all images.

**Table 1**. Geometric and optical parameters of the synthesized nanoparticles (length, aspect ratio, surface area to volume ratio), the normalized plasmon resonance bandwidth, $FWHM/\lambda_{PR}$, shift of the plasmon resonance after functionalization with NBT molecules (PR-shift), measured SERS signal intensity, $I_{1347}$, and the value proportional to the intensity per molecule, $I_{1347}/(S/V)$. The ratio of the absorption efficiency factor to the radius of the equivalent volume sphere, $Q_{abs}/R_{ev}$, determines the photothermal efficiency of the particles at the plasmon resonance wavelength. The average length and aspect ratio were determined from TEM data (top numbers) and from the plasmon resonance wavelength, assuming a constant average diameter of 19 nm (bottom numbers).

| Sample | $L_{TEM}$ <br> $L_{PR}$ <br> (nm) | $AR_{TEM}$ <br> $AR_{PR}$ | $S/V$ <br> (nm$^{-1}$) | $\frac{FWHM}{\lambda_{PR}}$ | PR-shift <br> (nm) | $I_{1347}$ <br> (a.u.) | $I_{1347}/(S/V)$ <br> $\times 10^{-3}$ <br> (a.u.) | $Q_{abs}/R_{ev}$ <br> (nm$^{-1}$) |
|---|---|---|---|---|---|---|---|---|
| NR 640 | 50±5 <br> 44 | 2.63±0.45 <br> 2.32 | 0.24 | 0.14 | 3 | 415 | 1.7 | 0.337 |
| NR 735 | 68±4.5 <br> 60 | 3.78±0.4 <br> 3.16 | 0.24 | 0.13 | 4 | 570 | 2.4 | 0.547 |
| NR 844 | 82±4.7 <br> 78 | 4.32±0.48 <br> 4.1 | 0.23 | 0.14 | 6 | 526 | 2.3 | 0.683 |
| NR 945 | 95±5.7 <br> 94 | 4.75±0.58 <br> 4.95 | 0.22 | 0.16 | 7 | 265 | 1.2 | 0.778 |
| BP 753 | 75±2.5 <br> 70 | 3±0.3 <br> 2.64 | 0.26 | 0.066 | 13 | 2152 | 8.3 | 0.996 |

In Table 1, the theoretical $AR_{PR}$ values were obtained from the equation relating the plasmon resonance wavelength of nanorods stabilized with a 3-nm CTAB layer [39]:

$$\lambda_{PR}(\text{nm}) = 113.3 \times AR + 396.7 . \quad (1)$$

The corresponding length $L_{PR}$ was obtained by multiplying this $AR_{PR}$ by the average rod diameter of 19 nm. Moreover, the average plasmon resonance bandwidth of the nanorods, expressed in terms of the standard parameter FWHM (Full Width at Half Maximum), also remains approximately constant (around 0.14). When comparing with the extinction spectrum of bipyramids (curve 5 in Fig. 1A), the remarkably high quality of their spectrum, with an FWHM parameter of 0.066—half that of all the nanorods—immediately stands out.

Further confirmation of the high quality of the bipyramids' extinction spectra comes from comparing experimental and theoretical extinction spectra (Fig. 2). For randomly oriented gold nanorods, extinction spectra were calculated using the T-matrix method. For the bipyramids, the longitudinal excitation spectrum is presented, calculated using MEM (practically coinciding with the COMSOL results).

From Fig. 2, it follows that the experimental nanorod spectra are significantly broader than the theoretical ones, as the latter do not account for particle-size (aspect-ratio) distribution. Quantitatively, this spectral broadening, expressed as the ratio of the experimental to the theoretical FWHM value, is 2.5 for the nanorods. In contrast, for the bipyramids, the same ratio is half that, approximately 1.25, even though the theoretical model also did not account for particle size distribution.

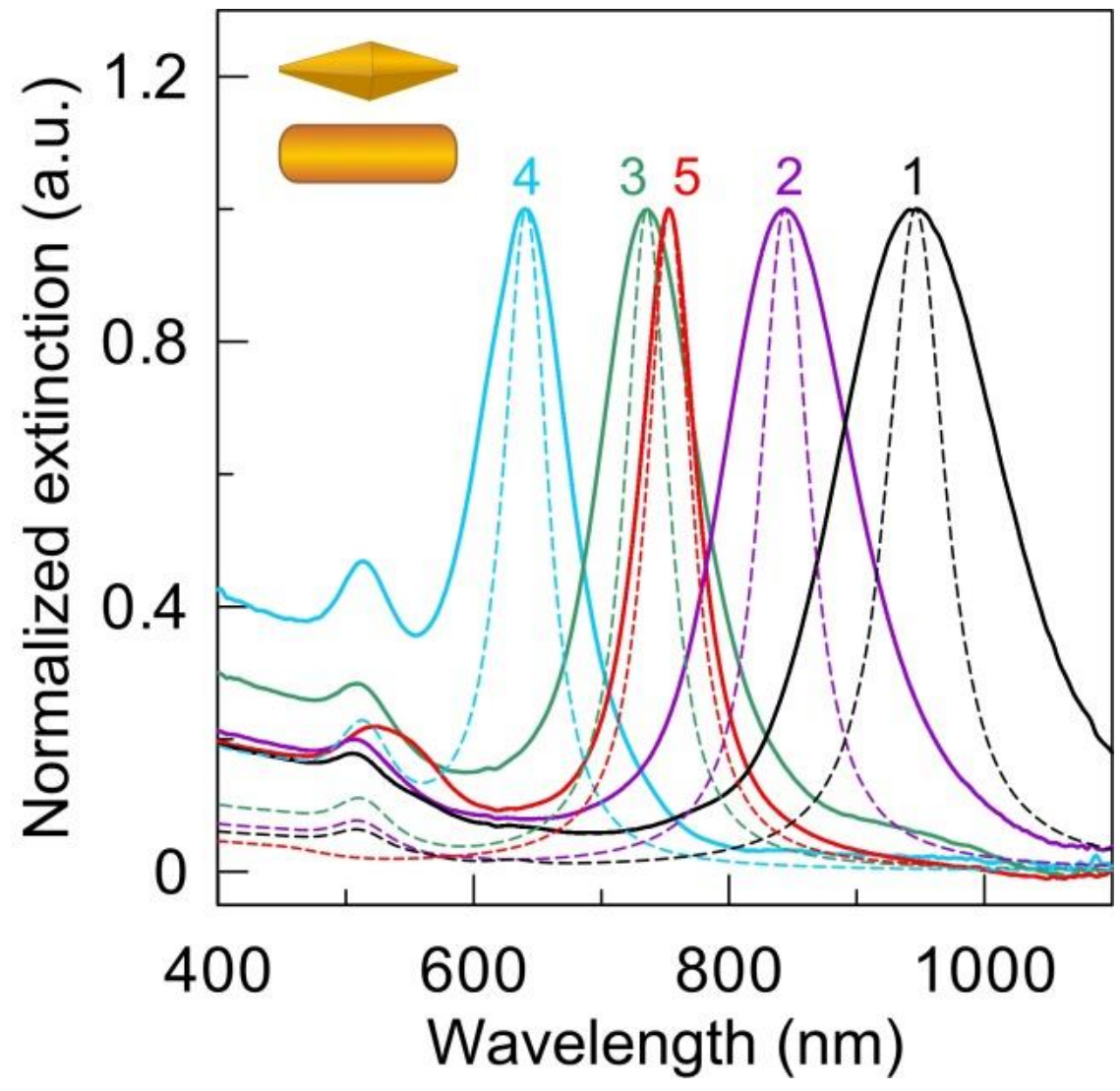

**Fig. 2.** Experimental and theoretical (dashed lines) normalized extinction spectra for the five samples listed in Table 1. The theoretical particle aspect ratios $AR_{PR}$ were used as fitting parameters to match the calculated and measured plasmon resonance wavelength.

After functionalization with the Raman reporter molecules NBT, the plasmon resonance maximum of all particles decreases in magnitude and shifts, as is typical, to the red region. It is clear that these changes are most pronounced in the bipyramids compared to the nanorods. Quantitatively, the plasmon resonance shift after functionalization for bipyramids is 2-4 times greater than that for nanorods. This is related to the higher analytical sensitivity of the bipyramids' plasmon resonance to the dielectric environment, which will be discussed below.

In addition to the usual geometric particle parameters $L$ and $AR$, Table 1 also provides the average surface area and the average surface area-to-volume ratio $S/V$. For nanorods with hemispherical ends, $S/V = 12AR/[d(3AR-1)]$, and for bipyramids modeled as inscribed bicones, $S/V = 6\sqrt{AR^2+1}/(d \cdot AR)$. As shown in Table 1, this parameter varies within a narrow range of 0.23–0.26. In the next section, we will use this parameter to estimate the SERS signal intensity per NBT reporter molecule.

The last column of Table 1 presents the theoretical value of $Q_{abs}/R_{ev}$, which is equal to the ratio of the absorption efficiency factor $Q_{abs}$ to the radius of the equivalent volume sphere $R_{ev}$. This parameter determines the photothermal efficiency of plasmonic nanoparticles per unit volume at a constant concentration $c(\text{g/mL})$, or, in other words, the photothermal efficiency per unit mass of nanoparticles. Indeed, the absorbed power per unit volume equals the product of the incident intensity $I(\text{W/cm}^2)$ and the numerical concentration $N(\text{cm}^{-3})$ multiplied by the absorption cross-section $C_{abs}(\text{cm}^2) = \pi R_{ev}^2 Q_{abs}$:

$$P_{abs}(\text{W/cm}^3) = I \times N \times C_{abs} = I \times \frac{c}{\frac{4}{3}\pi R_{ev}^3 \rho} \times \pi R_{ev}^2 Q_{abs} = I\,\frac{3c}{4\rho} \times \frac{Q_{abs}}{R_{ev}}, \quad (2)$$

where $\rho$ is the metal density. From equation (2), it follows that the absorbed power per unit volume is proportional to the ratio $Q_{abs} / R_{ev}$ at a constant mass-volume concentration of particles $c(\text{g/mL})$.

*3.2. Comparison of SERS spectra for nanorods and nanobipyramids*

Figure 3 shows the SERS spectra of functionalized gold nanorods and bipyramids. The most intense line at 1347 cm$^{-1}$ was used as a reference. Comparison of the spectra reveals a more intense peak for the functionalized bipyramids.

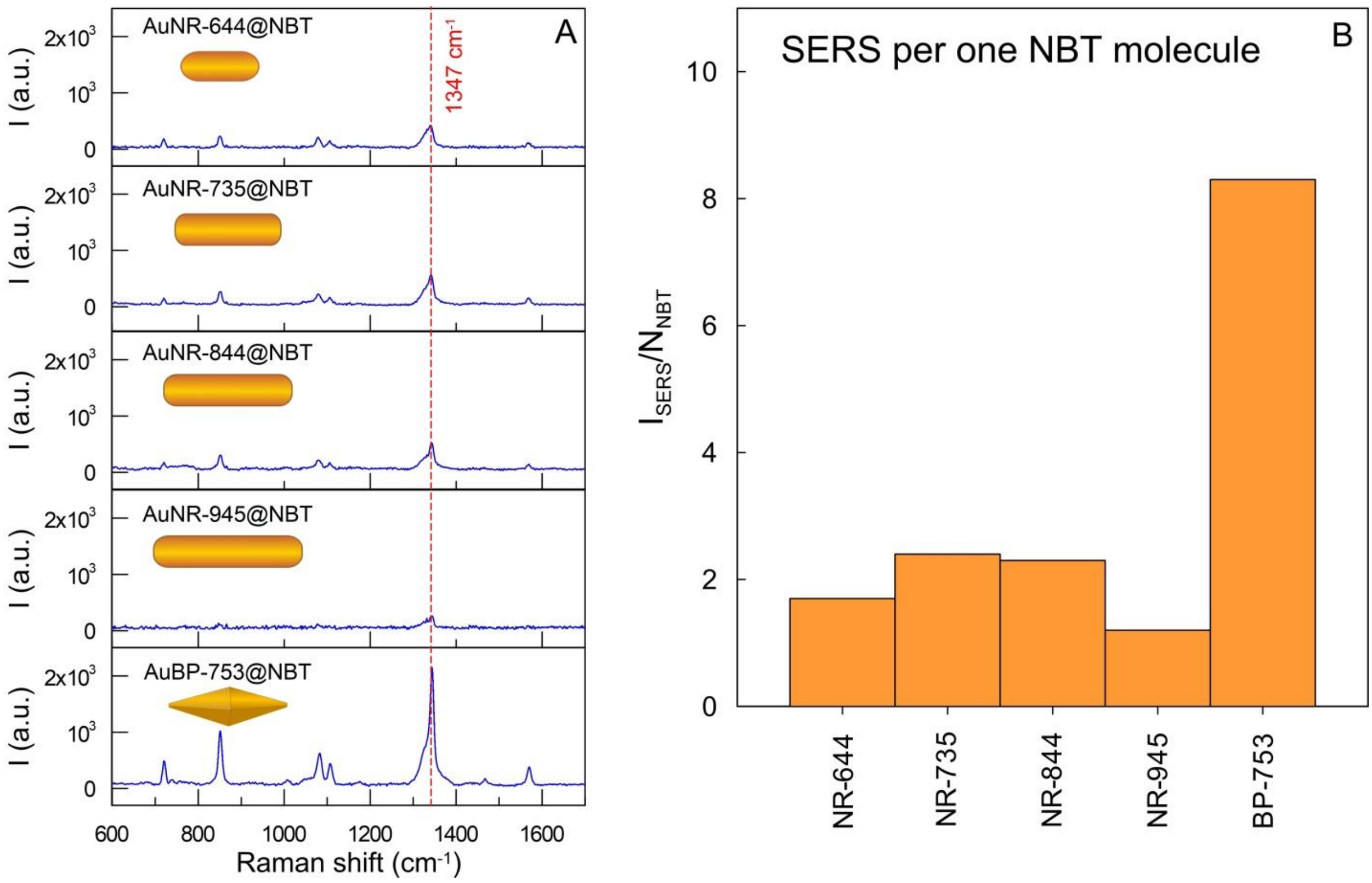

**Fig. 3**. SERS spectra measured for functionalized samples of nanorods and bipyramids (A). A red dashed line indicates the most intense characteristic spectral line at 1347 cm$^{-1}$. Panel B shows the intensities (arb. units) per one adsorbed NBT molecule. For bipyramids, this parameter is almost 4 times higher than that for nanorods.

To quantitatively assess the SERS efficiency of particles with different morphologies, theoretical estimates of the average fourth power of the local field amplitude near the particles are often used, or a slightly more rigorous estimate $\langle |E(\omega_L)|^2 |E(\omega_R)|^2 \rangle$ that accounts for the difference between the laser and Raman frequencies [58]. In this work, we use a simple estimate based on experimental extinction spectra, SERS spectra, and particle geometric parameters from TEM data. As mentioned earlier, the extinction and SERS spectra were recorded at approximately the same molar concentration of 0.1 mM or, correspondingly, at a mass-volume concentration of gold $19.7\times10^{-6}$ г/мл. For numerical estimates, we will assume that the number of molecules per particle is proportional to their surface area. Since functionalization was carried out with an excess of NBT, this assumption implies monolayer adsorption, and the number of molecules on a single particle is proportional to the ratio of the particle's surface area to the effective footprint area of one molecule $N_1 = S/S_1$. The total number of particles in 1 mL of colloid equals the ratio of the mass-volume concentration of gold to the mass of a single particle $N = m/m_1 = c/m_1$. The gold concentration was constant in all experiments (0.1 mM = $19.7\times10^{-6}$ г/мл), therefore, the number of reporter molecules in 1 mL of colloid is

$$N_R = N_1 \times N = \frac{S}{S_1}\frac{c}{m_1} = \frac{S}{S_1}\frac{c}{\rho V} = \left(\frac{c}{S_1\rho}\right)\frac{S}{V} = K\frac{S}{V}, \quad (3)$$

where, for our comparative estimates, the $K$ constant can be set to 1. Thus, under our experimental conditions, the number of molecules per particle is proportional to the ratio of its surface area to its volume.

Figure 3B presents the normalized SERS intensity $I/(S/V)$ values, which, as explained above, are proportional to the average SERS intensity from a single adsorbed NBT molecule. In this sense, this parameter provides a comprehensive, averaged characteristic of the efficiency of a plasmonic nanoparticle as an SERS platform. As shown in the figure, the SERS efficiency of bipyramids is approximately 4 times higher than that of nanorods.

*3.3. Analytical sensitivity of plasmon resonance in nanoparticles of different morphology to the dielectric environment*

The sensitivity of plasmon resonance (PR) to the dielectric environment is utilized for various analytical tasks aimed at determining target substances based on PR shifts [59]. Sensors based on this principle are somewhat similar to total internal reflection sensors, although the physical mechanism is entirely different in this case. In recent years, the most popular trend has been the development of fiber-optic sensors with a plasmonic nanostructure at the tip [29]. Theoretically, there are two approaches to estimating the analytical sensitivity of a PR sensor. In the most straightforward approach, the plasmonic response (e.g., the PR wavelength) is typically analyzed in a homogeneous dielectric medium with a varying refractive index (usually from 1.334 (water) to 1.6), and the sensitivity of different platforms is compared in terms of the derivative of the response with respect to the refractive index. For example, in the case of the PR wavelength, parameters such as $S_{PR} = \Delta\lambda_{PR} / \Delta n_m$ [60, 61] are compared. Within the dipole approximation, the polarizability of a particle of any shape and structure can be written as [53, 62]:

$$\alpha = \frac{3V}{4\pi} \frac{\varepsilon_{av} - \varepsilon_m}{\varepsilon_{av} + \varphi\varepsilon_m} = a_0{}^3 \frac{\varepsilon_{av} - \varepsilon_m}{\varepsilon_{av} + \varphi\varepsilon_m}, \quad (4)$$

where $\varepsilon_{av}$ is the average dielectric permittivity of particles with a complex structure, calculated using the DEM. The function $\varphi$ also depends on the shape and structure, and $\varepsilon_m$ is the dielectric permittivity of a homogeneous medium. Here are three simple examples:

1. A metal sphere with the dielectric permittivity $\varepsilon_{metal} = \varepsilon(\omega)$ in a homogeneous medium with dielectric permittivity $\varepsilon_m(\omega)$: $\varepsilon_{av} = \varepsilon(\omega)$, $\varphi = 2$.

2. A metal ellipsoid with the dielectric permittivity $\varepsilon_{metal} = \varepsilon(\omega)$ in a homogeneous medium $\varepsilon_m(\omega)$: $\varepsilon_{av} = \varepsilon(\omega)$, $\varphi = L_{a,b,c}^{-1} - 1$, where $L_{a,b,c}$ are the geometrical depolarization factors along the axes $a,b,c$, $\sum_{i=a,b,c} L_i = 1$. For a sphere $L_i = L = 1/3$.

3. A two-layered sphere with a metal core $\varepsilon_{metal} = \varepsilon_1(\omega)$, covered by a dielectric shell of thickness $s$ with a dielectric permittivity $\varepsilon_s = \varepsilon_2(\omega)$ in a homogeneous medium $\varepsilon_m(\omega)$:

$$\varepsilon_{av} = \varepsilon_2 \frac{1 + 2 f_{12}\alpha_{12}}{1 - f_{12}\alpha_{12}}, \quad \varphi = 2, \tag{5}$$

where $\alpha_{12} = (\varepsilon_1 - \varepsilon_2)/(\varepsilon_1 + 2\varepsilon_2)$, $f_{12} = a_1^3 / a_2^3 = a_1^3 / (a_1 + s)^3$. It follows from Eq. (4) that the PR resonance condition reads:

$$\mathrm{Re}(\varepsilon_{av}) = -\varphi\varepsilon_m . \tag{6}$$

In the Drude approximation, the metal dielectric function is given by

$$\varepsilon(\omega) = \varepsilon_{ib}(\omega) - \frac{\omega_p^2}{\omega(\omega + i\gamma_b)}, \tag{7}$$

where $\varepsilon_{ib}(\omega)$ accounts for the contribution of interband transitions, $\omega_p$ is the bulk metal plasma frequency (for gold, $\hbar\omega_p \simeq 9$ eV), and the damping constant $\gamma_b$ corresponds to a bulk sample. It does not account for additional damping mechanisms [62]. Combining (6) and (7), we find the plasmon resonance wavelength [53, 62]:

$$\lambda_{PR} = \lambda_p \left[\varepsilon_{ib} + \varepsilon_m \varphi\right]^{1/2} . \tag{8}$$

Using (8), one can obtain a theoretical estimate of the sensitivity of PR to changes in the refractive index of the external medium in the form of universal relations:

$$S_{PR} = \frac{\Delta\lambda_{PR}}{\Delta n_m} = \varphi \frac{\lambda_p n_m}{\sqrt{\varepsilon_{ib} + \varphi n_m^2}}, \tag{9}$$

$$\frac{\Delta\lambda_{PR}}{\lambda_{PR}} = \frac{\Delta n_m}{n_m}\left(1 - \frac{\lambda_p^2}{\lambda_{PR}^2}\varepsilon_{ib}\right), \tag{10}$$

where $\lambda_p = 2\pi\omega_p / c$ is the plasma wavelength (for gold, $\lambda_p \simeq 130\,\text{nm}$), and $c$ is the light velocity in vacuum.

Relation (9) explains the theoretically obtained [60, 61] and experimentally observed linear dependencies $\lambda_{PR} = f(n_m)$ with a constant coefficient $S_{PR}$, since $n_m$ usually changes within a narrow range of 1.3 to 1.5 (for water-glycerol mixtures). In this case, according to (10), the relative change $\Delta\lambda_{PR} / \lambda_{PR}$ turns out to be proportional to the relative change $\lambda_{PR} = f(n_m)$ with an accuracy of approximately a constant coefficient (for gold in water, it is approximately 0.5 [63]). Relations (9-10) also explain that the structure and shape of a particle decisively influence the angular slope of the dependence $\lambda_{PR} = f(n_m)$ through the parameter $\varphi$ in relation (9).

For the theoretical evaluation of the analytical sensitivity of a plasmonic sensor, one can also calculate the change in the plasmonic response resulting from the formation of a dielectric layer on the particle surface, for example, due to antigen-antibody interactions [64]. This formulation of the problem seems more justified from the perspective of modeling real experimental situations, where the process of analyte molecule adsorption on the surface of a plasmonic nanosensor is detected by the shift in plasmon resonance. This modeling process itself can be carried out in two variants: (1) analysis of the PR shift as a function of shell thickness with a constant refractive index; (2) analysis of the PR shift as a function of the shell's refractive index at its constant nanometer-scale thickness. The real situation is most likely a combination of these two variants, since as the thickness of the adsorbed biopolymer layer increases, the volume fraction of the aqueous buffer (or other solvent) will decrease. Below, we present comparative data for nanorods and BPs (bipyramids) in a homogeneous, infinite medium and in a shell of variable thickness with a constant refractive index.

Figure 4A shows the dependence of the plasmon resonance wavelength shift $\Delta\lambda_{PR}$ on the refractive index of the external medium $n_m$ for gold bipyramids and nanorods with a diameter of 15 nm and an aspect ratio of 3. For the bipyramids, calculations were performed using the 3D FEM COMSOL method under longitudinal excitation of the particles by an electric field. For the nanorods, the calculation was performed on randomly oriented nanoparticles using the T-matrix method. We also performed an approximate averaging over the orientations of the bipyramids and confirmed that the result differs only slightly from that presented in Fig. 4A.

From Fig. 4A, it follows that for nanoparticles with identical geometric parameters, the analytical sensitivity of bipyramids, measured by the plasmon resonance shift, is higher than that of nanorods. From a physical standpoint, this difference can be explained by the strong field localization near the vertices of the bipyramid (Fig. 4B) and the resulting more substantial change in the dipole moment upon variation of the external refractive index compared to nanorods (Fig. 4C). Qualitatively, the modulus of the local field near the bipyramid vertices is approximately an order of magnitude greater than the field near the ends of the nanorods. This also explains the aforementioned SERS signal data in Fig. 3.

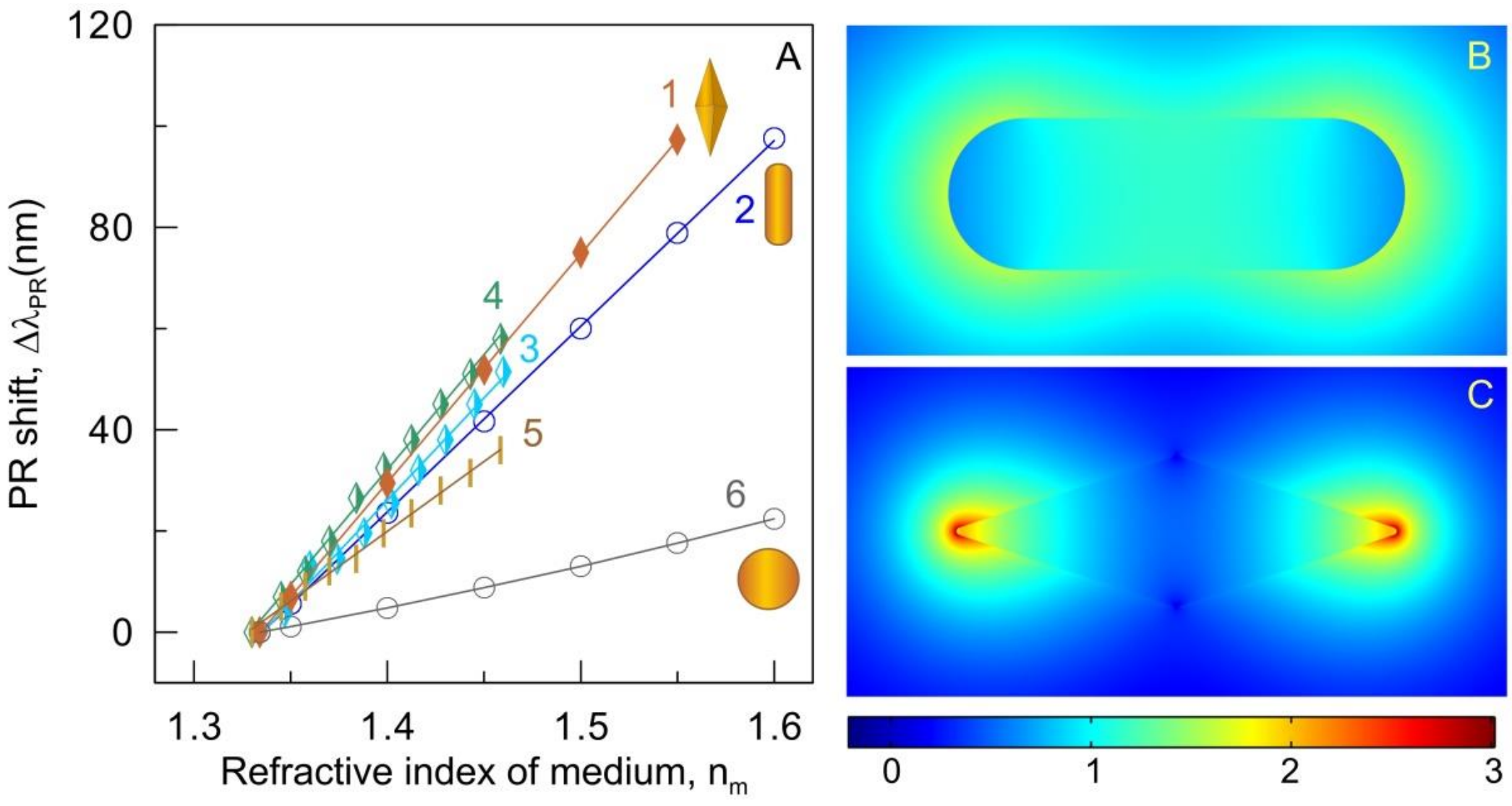

**Fig. 4**. Dependence of the plasmon resonance wavelength shift $\Delta\lambda_{PR}$ on the refractive index of the external medium ($n_m$). Calculation for gold BPs (1, COMSOL) and nanorods (2, T-

matrix method) with a diameter of 15 nm and an aspect ratio of 3. Experimental points 3 and 4 for gold BPs are reproduced from works by Chen et al. [28] ( $AR = 3.8$ ) and by Fang et al. [30] ( $AR = 3.3$ ), respectively. Experimental points (5) for gold nanorods with $AR = 3$ are reproduced from Fang et al. [30]. For comparison, the linear dependence (6) according to Mie theory for spheres with an equivolume diameter of 24 nm is also shown. Panels B and C show the distribution of the logarithm of the field modulus (log|E|/|$E_0$|) around the particle in water at the plasmon resonance wavelength for a nanorod (B) and a nano-bipyramid BP (C), respectively. The numbers on the color scale correspond to an order-of-magnitude change in the local field amplitude.

Additional confirmation of the strong influence of shape and local field is provided by line (6), obtained from Mie theory for spheres of equivalent diameter. For qualitative comparison with experimental data, we have reproduced in Fig. 3 data from the work of Chen *et al.* [28] for bipyramids with an aspect ratio of $AR = 3.8$ and from Fang et al. [30] for bipyramids with an aspect ratio of $AR = 3.3$. The last data are in close agreement with our simulations. It should also be noted that the particle sizes and tip rounding radii in the Chen *et al*. [28] samples were significantly larger than those used in our calculation for dependence (1); therefore, this comparison should be considered qualitative. This is precisely why the experimental line (4) for bipyramids with an aspect ratio of 3.8 lies below the theoretical line for an aspect ratio of 3 and the experimental line (4) by Fang *et al*. [30] for an aspect ratio of 3.3. The experimental data [30] for nanorods confirm a lower angular slope than that for BPs. In general, we conclude that the theoretical estimates of the angular slope $S_{PR}$ for BPs and nanorods are in qualitative agreement with the experimental data.

Now let's discuss the data for the case where the analyte forms a shell on the surface of a plasmonic nanoparticle. Figure 5A shows the dependence of the plasmon resonance wavelength shift $\Delta\lambda_{PR}$ on the shell thickness $s$ (curves 1, 2) or its volume fraction $g = V_s / (V_s + V_{Au})$ (curves 3, 4) for bipyramids, modeled as an inscribed bicone [39] with an aspect ratio of 3. For comparison, two particle diameters of 15 nm and 30 nm were chosen, which are close to the experimental conditions [39], with a typical refractive index of 1.5 for many biopolymers or silicon dioxide [65].

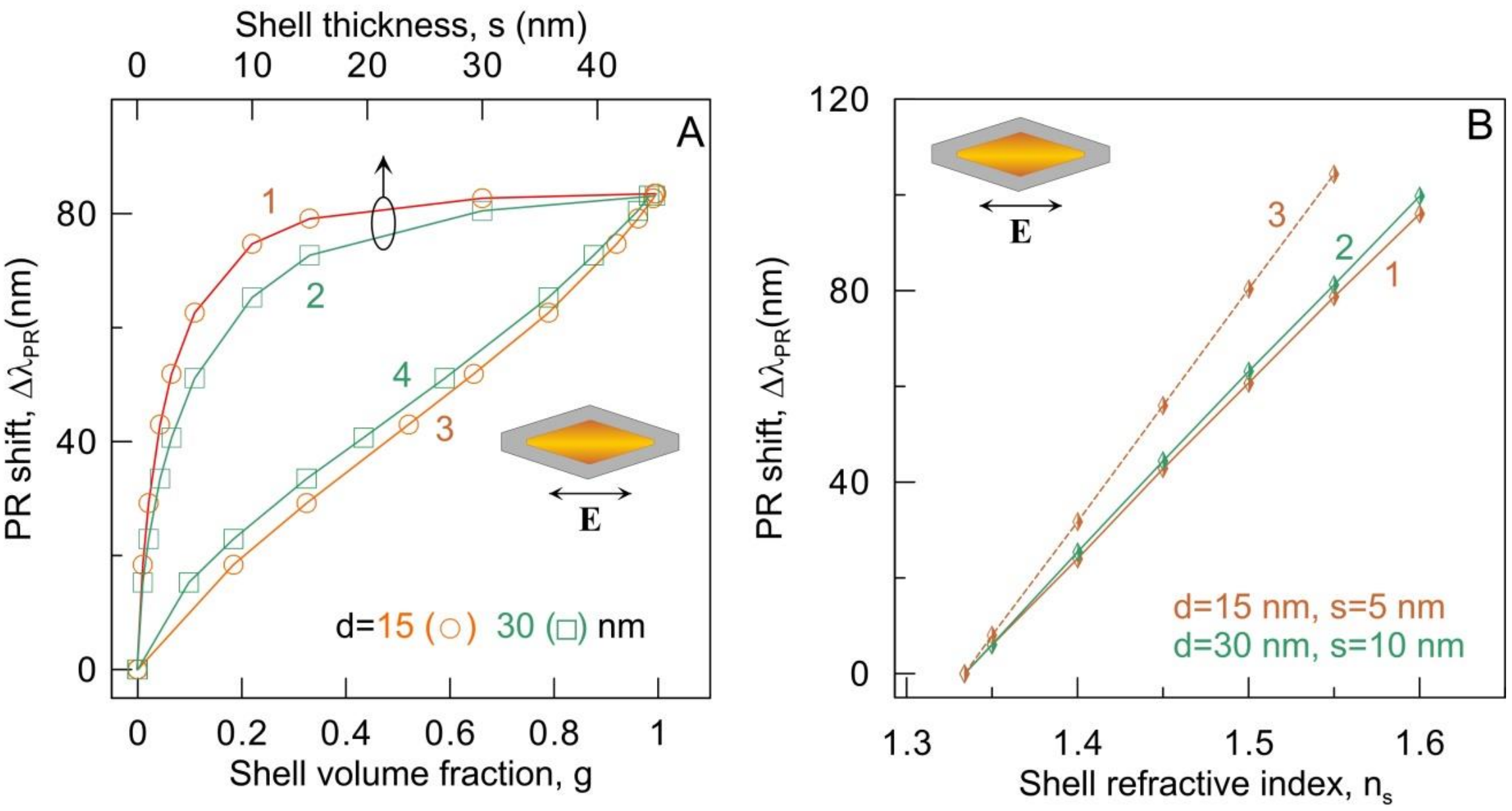

**Рис. 5**. (A) Dependence of the plasmon resonance wavelength shift $\Delta\lambda_{PR}$ on shell thickness (1, 2) or its volume fraction (3, 4). Calculation based on the analytical MEM theory for composite nanoparticles consisting of gold bicones with diameters of 15 nm (1, 3) and 30 nm (2, 4), an aspect ratio $AR = 3$, and a dielectric shell thickness $s$ ranging from zero to 45 nm. The refractive index of the shell is 1.5. (B) Dependence of the plasmon resonance wavelength shift $\Delta\lambda_{PR}$ of gold bicones on the refractive index of a dielectric shell with thicknesses of 5 nm (1) and 10 nm (2). For comparison, the dashed line (3) shows the dependence of the plasmon shift (1) for a particle in an infinite homogeneous medium. Particle diameters are 15 nm (1) and 30 nm (2), and the axial ratio is 3.

From plots 1 and 2, it can be seen that the main sharp plasmon resonance shift occurs when the shell thickness changes from 0 to 5-10 nm, at which point the shell volume fraction approaches 75-80%. In accordance with the field distribution shown in Figure 3C, this shell region precisely falls within the "hottest" area of the local electric field. Further increase in shell thickness has a lesser effect on the plasmon shift, which essentially becomes close to the PR-shift in an infinite dielectric medium with the shell's refractive index (cf. dependencies 1 and 3 in Figure 4B).

Figure 5B shows the dependencies of the plasmon resonance wavelength shift on the refractive index of the shell for two particle diameters, 15 nm and 30 nm, at shell thicknesses of 5 nm and 10 nm, which also differ by a factor of two. As shown, lines 1 and 2 are practically indistinguishable, which is explained by the approximately equal shell volume fractions in both nanocomposites. As mentioned above, increasing the shell thickness to that of an infinite homogeneous dielectric medium somewhat increases the angular slope of dependence (3). Still, it does not change the fundamental essence of the matter. Thus, the formation of dielectric analyte shells with a thickness of about 10 nm is essentially equivalent to the plasmon resonance shift when placed in an infinite analyte medium.

### *3.4. Comparison of thermoplasmonic properties for gold nanorods and nanobipyramids*

To compare the efficiency of converting laser radiation into heat, an experiment was conducted (Fig. 6A) comparing the heating of gold BPs and nanorods with equal extinction at 800 nm, a wavelength close to the plasmon resonance wavelength of both nanoparticle types. The concentration of nanorods was 0.062 mM (12 μg/mL), determined by absorbance at 400 nm. A 2 W laser with a wavelength of 800 nm was used for irradiation; the intensity of the expanded beam at the cuvette entrance was approximately 290 mW/cm². Irradiation was performed from above into a 1x1 cm cuvette containing 1 mL of suspension, so the optical path length from the top to the bottom of the cuvette was 1 cm. The temperature distribution (Fig. 6B) was recorded from the side using a thermal imager (Fig. 6A). The irradiation and temperature recording lasted about 6 minutes.

Figure 6C shows the dependencies of the temperature increment (relative to the water temperature under the same irradiation conditions) on the irradiation time. It can be seen that after 5-6 minutes of irradiation, the temperature increment reaches an approximately steady-state value of around 30 degrees. Although the heating kinetic curves are generally similar, we observe slightly higher efficiency for gold bipyramids, in agreement with the theoretical estimate of the $Q_{abs} / R_{ev}$ ratio in the last column of Table 1. Thus, theoretical and experimental data show that, despite the smaller metal volume per particle, the photothermal efficiency of bipyramids is of the same order (or even slightly higher) as that of widely used nanorods.

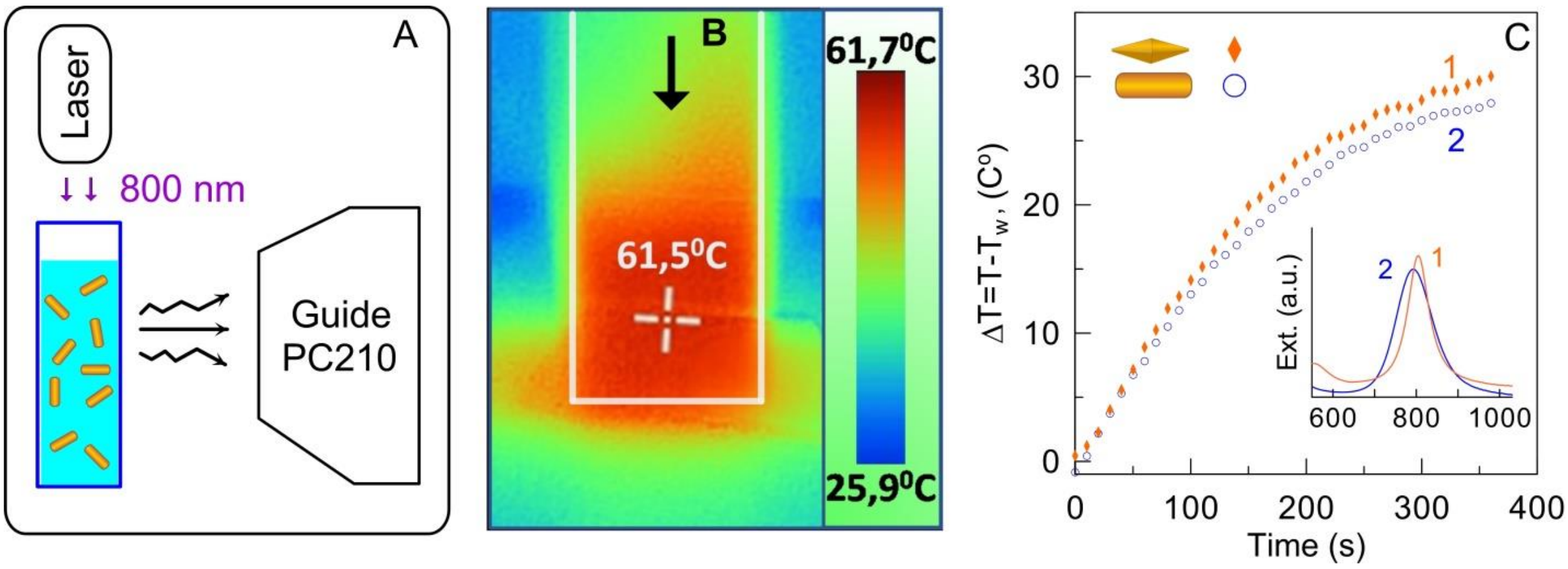

**Fig. 6**. Schematic of irradiation and recording (A), temperature distribution (B) in a cuvette with an aqueous suspension, and temporal dependencies of the temperature difference between nanobipyramid suspension (1) and nanorod suspension (2) and water temperature ( $\Delta T = T - T_w$) under identical irradiation conditions (C). The suspension volume in a 1×1 cm cuvette is 1 mL; the arrow in panel B indicates the direction of irradiation. The inset in panel C shows the extinction spectra of nanobipyramids (1) and nanorods (2) with equal extinction at 800 nm.

### *3.5. Comparison of the photothermal killing of E. coli bacteria using gold nanorods and nanobipyramids*

For PPT experiments with bacteria and nanoparticles, cells were mixed with nanoparticles in a 1:1 ratio, incubated for 15 minutes, and irradiated with a laser for 10 minutes. Four controls were used in the experiments: 1 – native bacteria without any treatment (B Cont), 2 – bacteria without particles irradiated with the laser (B-Part+L), 3 – bacteria with nanorods without laser irradiation (B+NR-L), 4 – bacteria with nanobipyramids without laser irradiation (B+BP-L). In the two positive experiments, 5 – (B+NR-L) and 6 – (B+BP-L), bacteria were incubated with nanoparticles for 15 min, then irradiated with a laser intensity of 290 mW/cm² for 10 min. All experiments were conducted using a 96-well plate.

Cell viability was assessed using the standard Alamar Blue test [56]. For this, 5 μL of resazurin was added to each 200 μL sample, and the mixture was incubated in a thermal chamber at 37°C for 1 hour. The results are presented in Fig. 7 and Table 2.

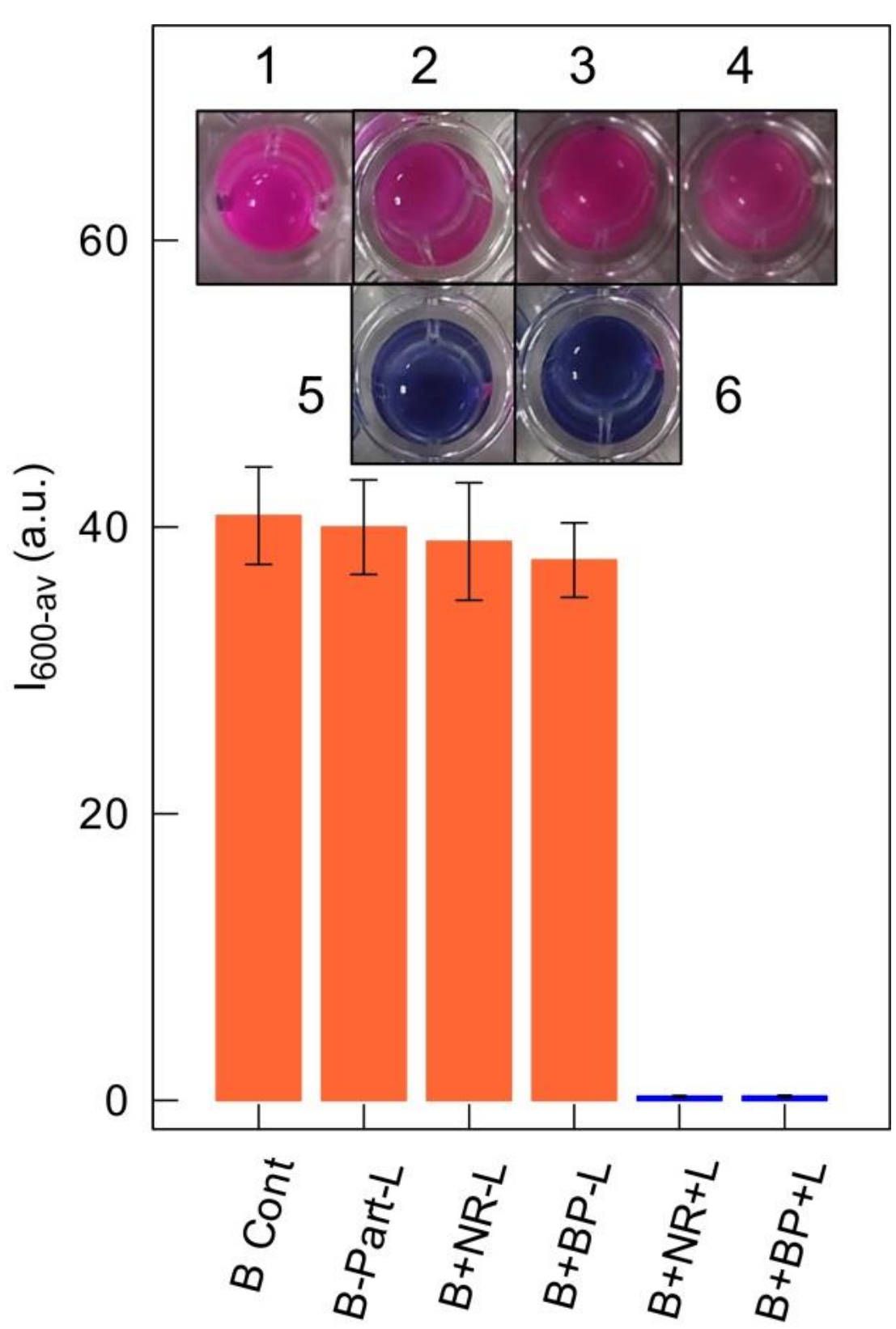

**Fig. 7**. Photograph of six plate wells after the Alamar Blue test. The numbers correspond to the designations of controls and experiments as per the text: 1 – B Cont, 2 – B-Part+L, 3 –

B+NR-L, 4 – B+BP-L, 5 – B+NR+L, 6 – B+BP+L. Below is a histogram of the average luminescence intensity in the 590-610 nm range with excitation at a wavelength of 530 nm.

From the presented data, it can be seen that both types of nanoparticles are effective thermosensitizers, causing complete bacterial death after 10 minutes of irradiation at a very low intensity of about 0.3 mW/cm², which is an order of magnitude lower than the intensities typically used for photothermal therapy (2 W/cm²).

**Table 2**. Fluorescence intensities of the samples at 600 nm and assessment of cell viability.

| Sample | Intensity $I_{600}$ (a.u.) | Cell viability, % |
|---|---|---|
| 1– B Control | 44.6 | 100 |
| 2– B-Part+L | 43.5 | 97.5 |
| 3– B+NR-L | 40.5 | 90.7 |
| 4– B+BP-L | 40.1 | 90 |
| 5– B+NR+L | 0.30 | 0.7 |
| 6– B+BP+L | 0.26 | 0.6 |

## 4. Conclusions

In this work, we have presented experimental and theoretical data comparing several important plasmonic parameters of the most popular anisotropic particles—gold nanorods—and less familiar, yet, in our view, quite promising plasmonic nanoparticles: gold bipyramids. Our goal was to draw researchers' attention to this new class of anisotropic plasmonic nanoparticles with a tunable plasmon resonance spanning 600-1200 nm. Due to their strong shape anisotropy, bipyramids generate a strong local field near their sharp tips when excited at their plasmon resonance wavelength. Such localization has important implications for potential biomedical applications. First, Raman reporter molecules located near these local-field "hot spots" emit a stronger SERS signal than conventional nanorods under the same conditions. Second, bipyramids exhibit greater analytical sensitivity to global or local dielectric environmental changes, leading to a larger plasmon resonance shift. Third, it can be expected that bipyramids will be promising nanoparticles as a plasmonic platform for bioimaging. Finally, thanks to their high absorption per unit mass of metal, bipyramids show promise as photothermal sensitizers that efficiently generate heat upon laser irradiation.

**Acknowledgment**

The study of the photothermal properties of bipyramids and plasmon resonance-mediated bacterial killing was supported by the Russian Science Foundation grant No. 24-65-00015, https://rscf.ru/project/24-65-00015. The synthesis of nanoparticles, their characterization, SERS investigations, and theoretical analysis were performed within the framework of the state assignment of the Ministry of Science and Higher Education of the Russian Federation for the Federal Research Center "Saratov Scientific Center of the Russian Academy of Sciences," theme No. 121031700141-6.

**Compliance with ethical standards**

This work does not involve studies with human participants or animals.

**Conflict of interest**

The authors of this work declare that they have no conflicts of interest.

**Declaration of generative AI and AI-assisted technologies in the manuscript preparation process.**

During the preparation of this work, the authors used DeepSeek v3 and Grammarly to check the English. After using these tools, the authors reviewed and edited the content as needed and take full responsibility for the content of the published article.

**Author contributions**:

Andrey Burov (nanoparticle synthesis, investigation, methodology, visualization, writing – review and editing), Sergey Zarkov (COMSOL simulations, software), Arina Drozd (PTT experiments), Igor Borisov (PTT and microbiology experiments), Elena Zavyalova (funding acquisition, project administration), Nikolai Khlebtsov (conceptualization, project administration, formal analysis, data curation, T-matrix simulations, visualization, writing – original draft, writing – review and editing). All authors have read and agreed to the published version of the manuscript.

## Highlights

- FWHM of gold nanobipyramids (AuBPs) is half of that for nanorods
- The SERS enhancement factor of AuBPs is three times higher than that of the nanorods
- PR shift of AuPBs is sensitive to variations in the external refractive index
- PR shift of AuPBs is sensitive to the dielectric shell parameters
- AuBPs effectively convert the light to heat and kill bacteria